\documentclass{vldb-modify}
\usepackage{graphicx}
\usepackage{balance}  

\usepackage{graphicx}
\usepackage{url}
\usepackage{color}
\usepackage{xcolor}
\PassOptionsToPackage{dvipsnames}{xcolor}
\usepackage{alltt}
\usepackage{hyperref} 
\hypersetup{
	citebordercolor={0 1 0},
	citecolor=blue,
	colorlinks=false,
	filebordercolor={0 .5 .5},
	filecolor=green,
	linkbordercolor={1 0 0},
	linkcolor=blue,
	menubordercolor={1 0 0},
	pageanchor=true,
	pagebackref=true,
	pdfborder={0 0 1},
	pdfpagelabels=true,
	pdftex,
	plainpages=false,
	urlbordercolor={0 1 1},
	urlcolor=blue,
}
\usepackage{syntax}
\usepackage{newfloat}
\usepackage{multicol}

\usepackage[labelfont=bf, skip=0pt]{caption}
\newcommand{\Comment}[1]{}

\newcommand{\SmallSpace}{\vspace*{-1.5ex}}

\newcommand{\myparatight}[1]{\smallskip\noindent{\bf {#1}.}}

\DeclareFloatingEnvironment[
fileext   = logr,
listname  = {List of Grammars},
name      = Grammar,
placement = !htbp
]{BNF}

\newcommand{\distance}{5pt}
\usepackage[ruled,vlined, noend]{algorithm2e}
\usepackage{mathtools}
\usepackage{url}
\usepackage{multirow}
\usepackage{mdwlist}
\usepackage{xspace}
\usepackage{amsmath}
\usepackage{amssymb}
\usepackage{misc}
\usepackage{subcaption}
\usepackage{listings}
\usepackage{enumerate}

\usepackage{enumitem}
\usepackage{adjustbox}
\usepackage{mdframed}

\definecolor{mygreen}{rgb}{0,0.6,0}
\definecolor{mygray}{rgb}{0.5,0.5,0.5}

\usepackage[capitalize,nameinlink]{cleveref}
\hypersetup{%
	bookmarksnumbered, bookmarksopen=true, bookmarksopenlevel=1,%
}

\crefname{figure}{Figure}{Figures}
\crefname{listing}{Query}{Queries}
\crefname{section}{Section}{Sections}
\crefname{table}{Table}{Tables}
\crefname{BNF}{Grammar}{Grammars}
\crefname{algorithm}{Algorithm}{Algorithms}

\newcommand{\incode}[1]{\lstinline{#1}}

\newcommand{\eat}[1]{}

\newcommand{\dsl}{\textsc{Aiql}\xspace}

\newif \ifcomments
\commentstrue

\ifcomments
    \newcommand{\kjee}[1]{{-\textcolor{red}{#1}-}}
\else
    \newcommand{\kjee}[1]{}
\fi

\newcommand{\specialcell}[2][c]{%
  \begin{tabular}[#1]{@{}l@{}}#2\end{tabular}}
\newcommand{\eg}{e.g., }
\newcommand{\ie}{i.e., }

\SetKwInput{KwInput}{Input}
\SetKwInput{KwOutput}{Output}
\SetKwProg{Fn}{Function}{}{end}

\SetCommentSty{mycommfont}

\begin{document}
\date{}


\title{A Query System for Efficiently Investigating Complex Attack Behaviors for Enterprise Security}
%

%

\author{
\alignauthor
Peng Gao$^1$\; Xusheng Xiao$^2$\; Zhichun Li$^3$\; Kangkook Jee$^3$\; Fengyuan Xu$^4$\; Sanjeev R. Kulkarni$^5$\; Prateek Mittal$^5$\\
       \affaddr{$^1$UC Berkeley\; $^2$Case Western Reserve University\; $^3$NEC Laboratories America, Inc}\;
 $^4$Nanjing University\; $^5$Princeton University\\
       \email{\normalsize penggao@berkeley.edu\; xusheng.xiao@case.edu\; \{zhichun,kjee\}@nec-labs.com\; fengyuan.xu@nju.edu.cn \{kulkarni,pmittal\}@princeton.edu}
}

%
%
%
%
%
%
%

%


\maketitle

\thispagestyle{empty}
\pagestyle{empty}

\begin{abstract}
The need for countering Advanced Persistent Threat (APT) attacks has led to the solutions that ubiquitously monitor system activities in each enterprise host, and perform timely attack investigation over the monitoring data for uncovering the attack sequence.
However, existing general-purpose query systems 
lack explicit language constructs for expressing key properties of major attack behaviors, and their semantics-agnostic design often produces inefficient execution plans for queries.
%
To address these limitations, we build \dsl, a novel query system that is designed with novel types of domain-specific optimizations to enable efficient attack investigation.
\dsl provides (1) domain-specific data model and storage for storing the massive system monitoring data, (2) a domain-specific query language, \emph{Attack Investigation Query Language (\dsl)} that integrates critical primitives for 
expressing major attack behaviors, and (3) an optimized query engine based on the characteristics of the data and 
the semantics of 
the query to efficiently schedule the execution.
We have deployed \dsl in NEC Labs America comprising 150 hosts. 
In our demo, we aim to show the complete usage scenario of \dsl by (1) performing an APT attack in a controlled environment, 
and (2) using \dsl to investigate such attack by querying the collected system monitoring data that contains the attack traces.
The audience will have the option to perform the APT attack themselves 
under our guidance, and interact with the system and investigate the attack via issuing queries and checking the query results through our web UI.

\end{abstract}



\section{Introduction}
\label{sec:intro}



Advanced Persistent Threat (APT) attacks are sophisticated (involving many individual attack steps across many hosts and exploiting various vulnerabilities) and stealthy (each individual step is not suspicious enough), plaguing many well-protected businesses with significant 
losses~\cite{target, equifax}.
In order for enterprises to counter APT attacks, recent approaches based on \emph{ubiquitous system monitoring} have emerged as an important solution for \emph{monitoring system activities and performing attack investigation}~\cite{backtracking,reduction,aiql,saql}.
System monitoring observes system calls at the kernel level to collect system-level events
that record system interactions among system entities (\eg processes, files, and network sockets).
Collection of system monitoring data enables security analysts to investigate these attacks by \emph{querying attack behaviors} over the historical data.

\eat{
\begin{figure*}[!ht]
	\centering
	\includegraphics[width=0.9\textwidth]{figs/moti-crop.pdf}
	\caption{Major types of attack behaviors (\cref{case:c5:comp,inv:d2,case:anomaly} show the corresponding \dsl queries)}
	\label{fig:moti}
\end{figure*}
}

\begin{figure*}[!ht]
	\centering
	\includegraphics[width=0.9\textwidth]{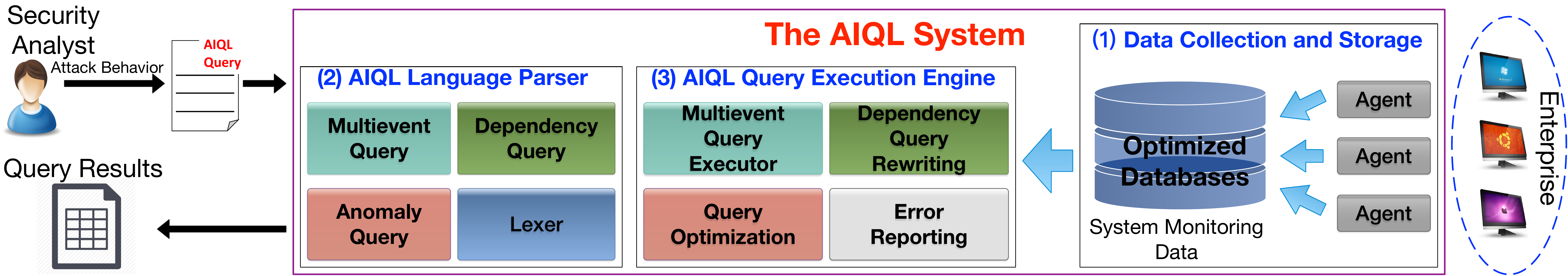}
	\caption{Architecture of the \dsl system}
	\label{fig:architecture}
\end{figure*}

Attack investigation is a time-sensitive task. However, there are two major challenges for building a query system to support efficient and timely attack investigation:

\emph{(1) Attack Behavior Specification}: 
The system needs to provide a query language with specialized constructs for expressing major 
attack behaviors:
\textbf{a. Multi-step attacks:} complex attacks such as APTs typically involve multiple system activities that are 
connected by specific attribute relationships (\eg the same process reads a sensitive file and accesses the network) or temporal relationships (\eg file read happens before network access), which requires language constructs to easily specify \emph{relationships among activities};
\textbf{b. Dependency tracking of attacks}: 
dependency tracking
is widely used in investigation to track causality of data for discovering the attack entry~\cite{backtracking}, which requires language constructs to easily \emph{chain constraints among activities};
\textbf{c. Abnormal system behaviors}:
frequency-based anomaly models are widely used to investigate abnormal 
system behaviors, such as network access spikes,
which requires language constructs to easily specify \emph{sliding windows} and \emph{statistical aggregation} of system activities and compare the aggregate results
in the current window with the results in previous windows.


\emph{(2) Timely Big-Data Analysis}: 
System monitoring produces a huge amount of daily logs~\cite{backtracking,reduction} ($\sim$ 50 GB per day for 100 hosts),
and the investigation of APT attacks typically requires the enterprise to keep at least a $0.5\sim 1$ year worth of data.
Such \emph{a huge amount of data} poses challenges for a timely investigation: the system needs to provide efficient data storage and query execution engine.

Unfortunately, existing query systems do not address both of these inherent challenges:
(1) Existing query languages in relational databases (e.g., PostgreSQL),
graph databases (e.g., Neo4j),
and other NoSQL databases (e.g., MongoDB, SPARQL)
lack explicit constructs to chain constraints among system activities and specify their relationships. 
To specify 
an attack behavior with multiple steps, 
these languages often lead to  large queries with many joins and constraints mixed together, posing great challenges for performance tuning. 
Constructing such queries correctly is also time consuming and error-prone.
Moreover, none of these languages provide explicit constructs for expressing behavioral models with accesses to historical aggregate results;
(2) System monitoring data is generated with a timestamp on a specific host in the enterprise, exhibiting strong \emph{spatial and temporal properties}. 
However, existing query systems are designed to work with general-purpose data thus missing opportunities for optimizations based on the domain data characteristics,
which might lead to some queries executing very inefficiently.

To address these challenges,
we build \dsl~\cite{aiql},
a system that enables security analysts to perform efficient attack investigation via querying system monitoring data.
\dsl employs three novel types of optimizations:
(1) \dsl provides a domain-specific query language, \emph{Attack Investigation Query Language (\dsl)}, which is optimized 
to express the three aforementioned types of attack behaviors;
(2) \dsl provides domain-specific \emph{data model and storage} for scaling the storage;
(3)\dsl optimizes the query engine based on the domain-specific characteristics of the system monitoring data and the semantics of the query
for efficient execution.

We have deployed the \dsl system in NEC Labs America comprising 150 hosts and made a demo video~\cite{aiqldemovideo}.
The system has been selected as part of commercialization process and integrated in the NEC Corporation's security intelligence solution, which won the first place in the 2016 CEATEC 
Award~\cite{ceatec16}.
In our demo, we aim to show the complete usage scenario of \dsl.
To achieve this goal, we first perform an APT attack in a controlled environment (for protecting the normal business) that exfiltrates sensitive data from database server by exploiting multiple vulnerabilities in multiple steps. 
The system monitoring data that contains the attack traces is collected by our data collection agents and stored in our optimized databases. 
Then, we use \dsl to investigate the attack by querying the collected data.
The audience will have the option to perform the APT attack themselves under our guidance, 
and interact with the system and investigate the attack via issuing queries and checking the query results through our web UI.  
The audience will also experience the superiority of \dsl by comparing the conciseness and performance of \dsl queries with SQL queries executed in PostgreSQL databases.

\eat{

In the demo, we will use \dsl to investigate an APT attack conducted by our white hat hackers in an isolated enterprise environment.
The audience will
have the option to  
interact with the system and investigate the attack via issuing queries and checking 
the query results through a web UI.
The audience will also experience the superiority of \dsl by comparing the conciseness and performance of \dsl queries with SQL queries.
}


%

\section{The AIQL System Architecture}
\label{sec:overview}

\cref{fig:architecture} shows the architecture of the \dsl system.
\dsl takes an input query from the user (\eg security analyst) that specifies certain attack behaviors to be investigated, executes the query, and retrieves the matched results.
\eat{
\dsl consists of three components:
(1) the data collection agents are deployed across enterprise hosts and collect information about system calls from kernels.
The collected system monitoring data is sent to the central server and stored in our optimized 
data storage;
(2) the language parser parses the input query;
(3) the query execution engine executes the query
to search for the desired attack behaviors
}

\subsection{Data Collection and Storage}
\label{subsec:collection-storage}

\myparatight{Data Model}
System monitoring data records the interactions
among 
system entities as system events.
Each of the recorded event occurs on a particular host at a particular time, thus exhibiting strong spatial and temporal properties.
In our data model, we consider \emph{system entities} as files, processes, and network connections.
We consider a \emph{system event} as the interaction between two system entities represented 
as \emph{$\langle$subject, operation, object$\rangle$} (SVO).
Subjects are processes originating from software applications (\eg Firefox), and objects can be files, processes, and network connections.
We categorize system events into three types according to their objects, namely \emph{file events}, \emph{process events}, and \emph{network events}.

\myparatight{Data Collection}
We develop data collection agents based on mature system monitoring frameworks: auditd
for Linux, ETW
for Windows, and DTrace
for MacOS.
Our agents are deployed across servers, desktops, and laptops in the enterprise and collect critical security-related attributes
(\eg file name, process executable name, IP, port, etc.; details in \cite{aiql}).

\eat{
\begin{table}[t]
	\centering
	\caption{Representative attributes of system entities}\label{tab:entity-attributes}
	\begin{adjustbox}{width=0.4\textwidth}
		\begin{tabular}{|l|l|}
			\hline
			\textbf{Entity}		&\textbf{Attributes}\\\hline
			File				&Name, Owner/Group, VolID, DataID, etc.\\\hline
			Process			&PID, Name, User, Cmd, Binary Signature, etc.\\\hline
			Network Connection	& IP, Port, Protocol \\\hline
		\end{tabular}
	\end{adjustbox}
	
\end{table}

\begin{table}[t]
	\centering
	\caption{Representative attributes of system events}\label{tab:event-attributes}
	\begin{adjustbox}{width=0.4\textwidth}
		\begin{tabular}{|l|l|}
			\hline
			\textbf{Operation}		& Read/Write, Execute, Start/End, Rename/Delete\\\hline
			\textbf{Time/Sequence}		& Start Time/End Time, Event Sequence\\\hline
			\textbf{Misc.}		& Subject ID, Object ID, Failure Code\\\hline
		\end{tabular}
	\end{adjustbox}
	
\end{table}
}

\myparatight{Data Storage}
Querying complex attack behaviors typically requires the efficient support for joins. 
Compared to graph databases and other NoSQL databases, relational databases come with mature indexing mechanisms and 
are more scalable to the massive data in our context.
Thus, in \dsl, we store the collected system monitoring data in relational databases (PostgreSQL and
Greenplum).
We further optimize the write throughput and the data storage using techniques such as data deduplication and in-memory indexes, batch commit, time and space partitioning, and hypertable (details in \cite{aiql}).

\subsection{AIQL Query Language}
\label{subsec:collection-storage}


We build the \dsl language using ANTLR 4.
Our language uniquely integrates a series of critical primitives for concisely expressing three major types of attack behaviors.

\subsubsection{Multievent AIQL query}
\label{subsubsec:multievent}

\dsl provides explicit 
 constructs for system events, spatial/temporal constraints, and event temporal/attribute relationships, which facilitates the specification of multi-step attack behaviors.
\cref{case:c5:comp} shows a multievent \dsl query that investigates the data exfiltration from database server:
the attacker leverages OSQL utility (\incode{osql.exe}) to dump the database content (\incode{backup1.dmp}) and 
runs a malware (\incode{sbblv.exe}) to send the dump back to his host (\incode{XXX.129}).
Four event patterns are declared (Lines 3-6) with two global constraints (Lines 1-2), a temporal relationship (Line 7), and an implicit attribute relationship (Lines 4-5 specify the same \incode{f1} in both events).
Desired attributes of matched events are returned (Line 8) with context-aware syntax shortcuts adopted (\ie \incode{p1} $\rightarrow$ \incode{p1.exe_name}, \incode{f1} $\rightarrow$ \incode{f1.name}, \incode{i1} $\rightarrow$ \incode{i1.dst_ip}).

\begin{lstlisting}[captionpos=b, caption={Data exfiltration from database server}, label={case:c5:comp}]
(at "mm/dd/2018") // time window (obfuscated)
agentid = xxx // SQL database server (obfuscated)
proc p1["%cmd.exe"] start proc p2["%osql.exe"] as evt1
proc p3["%sqlservr.exe"] write file f1["%backup1.dmp"] as evt2
proc p4["%sbblv.exe"] read file f1 as evt3
proc p4 read || write ip i1[dstip="XXX.129"] as evt4
with evt1 before evt2, evt2 before evt3, evt3 before evt4
return distinct p1, p2, p3, f1, p4, i1
\end{lstlisting}
\vspace{-1ex}

\subsubsection{Dependency AIQL query}
\label{subsubsec:dependency}

\dsl provides explicit 
constructs for chaining constraints among system events in the form of event path, which facilitates the dependency tracking of attacks.
\cref{inv:d2} shows a forward dependency \dsl query that investigates the ramification of a malware (\incode{info_stealer}), which originates from Host 1 (\incode{agentid = 1}) and affects Host 2 (\incode{agentid = 2}) through an Apache web server. 
An example execution result may show that \incode{p3} is the \incode{wget} process that downloads the malicious script from Host 2.
The \incode{forward} keyword (Line 2) specifies the temporal order of the events: left event occurs earlier.
The operation \incode{connect} (Line 4) indicates that the tracking is across different hosts.


\begin{lstlisting}[captionpos=b, caption={Forward tracking for malware ramification}, label={inv:d2}]
(at "mm/dd/2018") // time window (obfuscated)
forward: proc p1["%/bin/cp%", agentid = 1] ->[write] file f1["/var/www/%info_stealer%"] 
<-[read] proc p2["%apache%"] 
->[connect] proc p3[agentid=2] // tracking across hosts
->[write] file f2["%info_stealer%"] 
return f1, p1, p2, p3, f2 
\end{lstlisting}
\vspace{-1ex}

\subsubsection{Anomaly AIQL query}
\label{subsubsec:anomaly}


\dsl provides explicit 
constructs for sliding windows, aggregation functions, and accesses to historical aggregate results, which facilitates the specification of frequency-based anomaly models.
\cref{case:anomaly} shows an anomaly \dsl query that specifies a 1-minute sliding window (Line 3) and computes a moving average (Line 7) to investigate processes on the database server (Line 2) that transfer a large amount of data to a suspicious IP (\incode{XXX.129}).
An example execution result may show that the process \incode{p} is \incode{sbblv.exe}, which is suspicious and deserves further investigation.

\begin{lstlisting}[captionpos=b, caption={Large data transfer from database server}, label={case:anomaly}]
(at "mm/dd/2018") // time window (obfuscated)
agentid = xxx // SQL database server (obfuscated)
window = 1 min, step = 10 sec
proc p write ip i[dstip="XXX.129"] as evt
return p, avg(evt.amount) as amt
group by p
having (amt > 2 * (amt + amt[1] + amt[2]) / 3)
\end{lstlisting}
\vspace{-2ex}

\begin{figure}[t]
	\centering
	\includegraphics[width=0.41\textwidth]{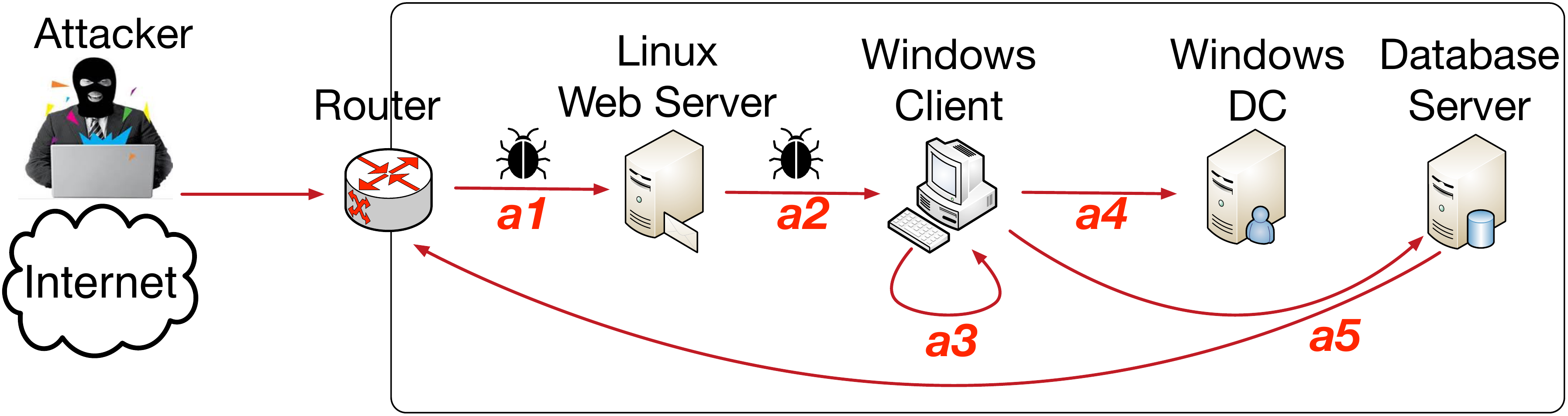}
	\caption{Demonstration setup for the APT attack}
	\label{fig:apt}
	\vspace*{-1ex}
\end{figure}

\subsection{AIQL Query Execution Engine}
\label{subsec:engine}


Our query execution engine leverages domain-specific characteristics of the data and the semantics of the query to efficiently schedule the execution.
Optimizing a query with many constraints is a difficult task due to the complexities of joins and constraints.
For a multievent query, 
\dsl addresses this challenge by synthesizing a \emph{SQL data query} for \emph{every} event pattern and schedules the execution of these data queries using our \emph{optimized scheduling strategy}, rather than weaving all the joins and constraints together in a large SQL query and relying on the inefficient default SQL engine scheduling.
Our \emph{optimized scheduling strategy} (details in~\cite{aiql}) has two key insights:
(1)
for a query with multiple event patterns, we prioritize the search of event patterns with higher pruning power, maximizing the reduction of irrelevant events as early as possible;
(2)
we partition the query into independent sub-queries along the temporal (\ie time window) and spatial (\ie agent ID) dimensions and execute these sub-queries in parallel.
For a dependency query, the parser compiles it to a semantically equivalent multievent query for execution.
For an anomaly query, the engine partitions the events into sliding windows by the timestamp, computes the aggregate results, and enforces the filters.

\section{Demonstration Outline}
\label{sec:demo}

\myparatight{Demonstration Setup}
We have deployed \dsl in NEC Labs America comprising 150 hosts.
The purpose of our demo is to illustrate the complete usage scenario of \dsl and showcase its superiority in enabling efficient attack investigation.
To achieve this goal, in our demo, we perform an APT attack in a controlled environment (\cref{fig:apt}) using a set of known exploits. The APT attack consists of five steps as follows:

\begin{itemize}[label={\arabic*.}, noitemsep, topsep=1pt, partopsep=1pt, listparindent=\parindent, leftmargin=*]
	\item[\emph{a1}] \emph{Initial Compromise}: The attacker first exploits the UnReal IRC server remote code execution vulnerability~\cite{cve1short} to create a telnet connection to his host.
	
	\item[\emph{a2}] \emph{Malware Infection}: The attacker uploads a malware via the connection and waits for the malware to infect other hosts to gain access to the intranet.

	\item[\emph{a3}] \emph{Privilege Escalation}: With the access to the intranet, the attacker leverages other vulnerabilities~\cite{cve2short} to escalate his privilege and executes memory dumping tools (Mimikatz, Kiwi) to obtain administrator credentials.
	
	\item[\emph{a4}] \emph{Obtain User Credentials}: The attacker 
	penetrates into the domain controller and executes password dumping tools (PwDump7.exe, WCE.exe) to obtain the credentials of all users.
	
	
	\item[\emph{a5}] \emph{Data Exfiltration}: Finally, the attacker penetrates into the database server
	and dumps the 
	data back to his host.
\end{itemize}

\myparatight{Live End-to-End Investigation Procedure}
%
After performing the attack, the system monitoring data that contains the attack traces is collected by our data collection agents and stored in our optimized databases.
Next, we begin the attack investigation process by constructing and iteratively revising \dsl queries.
Assuming no prior knowledge of the attack,
we start the investigation by first constructing an anomaly \dsl query and
identify a process ``powershell.exe'' transferring large data to a suspicious external IP ``XXX.129'' from the database server.
We then construct a multievent \dsl query to investigate the files read by this process and identify a database dump file ``db.bak''.
We further investigate the creation process of this dump file and identify ``sqlservr.exe'', which is a standard SQL server process with verified signature.
We also confirm that the process ``powershell.exe'' creates a connection to the IP ``XXX.129'' before the data transfer.
This confirms the existence of data exfiltration from the database server and completes the investigation of the step \emph{a5}.
We follow a similar procedure for investigating the steps \emph{a1-a4}.
Please refer to 
\cite{aiqldemoweb} for more investigation details and all \dsl queries.
%

\begin{figure}[!ht]
	\centering
	\includegraphics[width=0.41\textwidth]{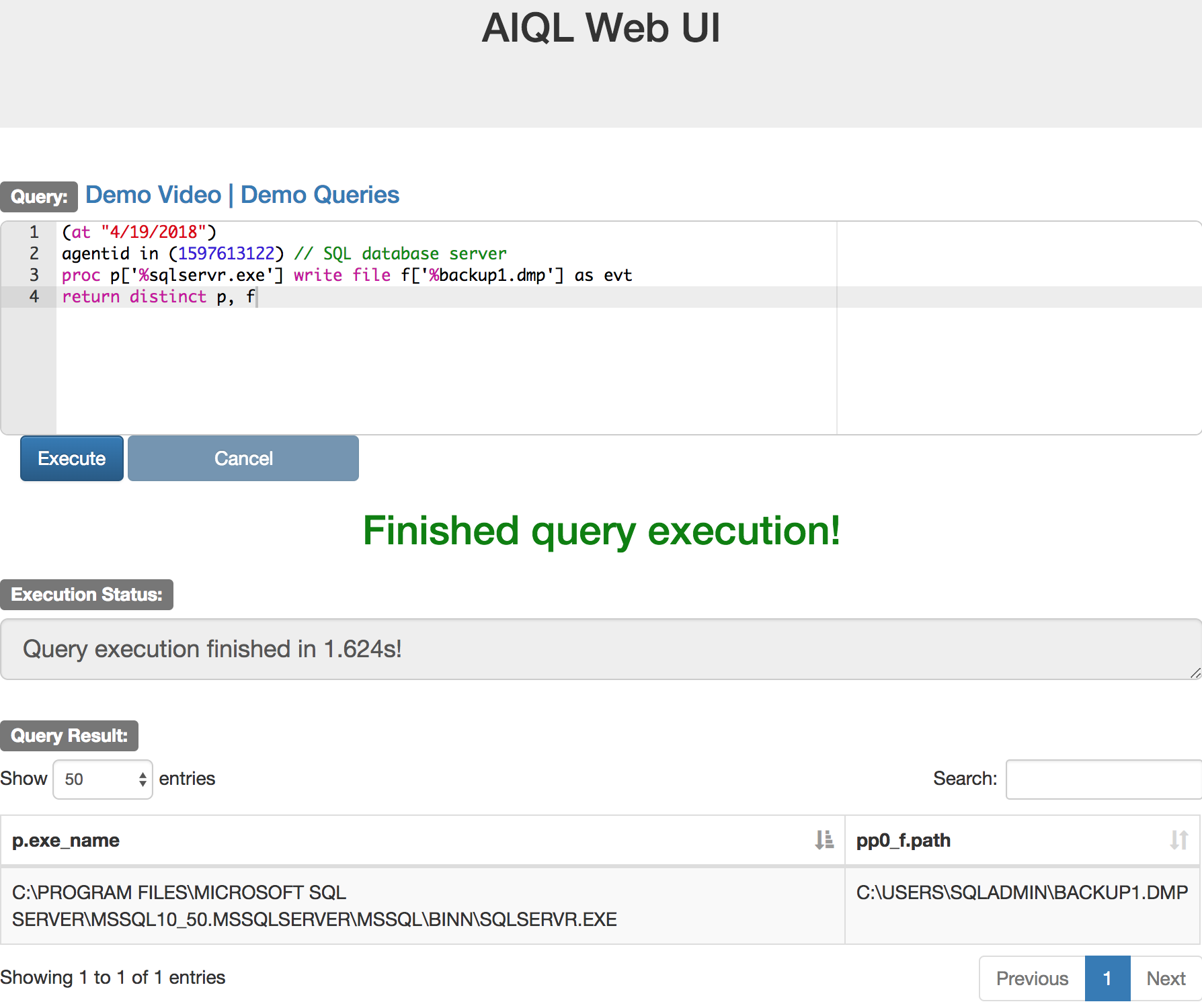}
	\caption{Web UI of the \dsl system}
	\label{fig:ui}
\end{figure}

\myparatight{Web UI}
In our demo, the audience will have the option to perform the attack under our guidance and do the investigation themselves by interacting with \dsl.
To facilitate such interaction, we build a web UI (\cref{fig:ui}) upon Apache Tomcat.
Our web UI consists of 
(1) an input box for entering \dsl queries, 
(2) an execution status area to show the query execution time, and 
(3) an interactive table that visualizes and manages the execution results.
Furthermore, our web UI provides query editing and result analysis features to facilitate efficient investigation:
(1) syntax highlighting for query construction,
(2) syntax checking for query debugging,
and
(3) sorting and searching for result management.
To get a better sense of how to use the web UI and interpret the query results, please refer to our demo video~\cite{aiqldemovideo}.


\begin{figure}[!tp]
\centering
\includegraphics[width=0.41\textwidth]{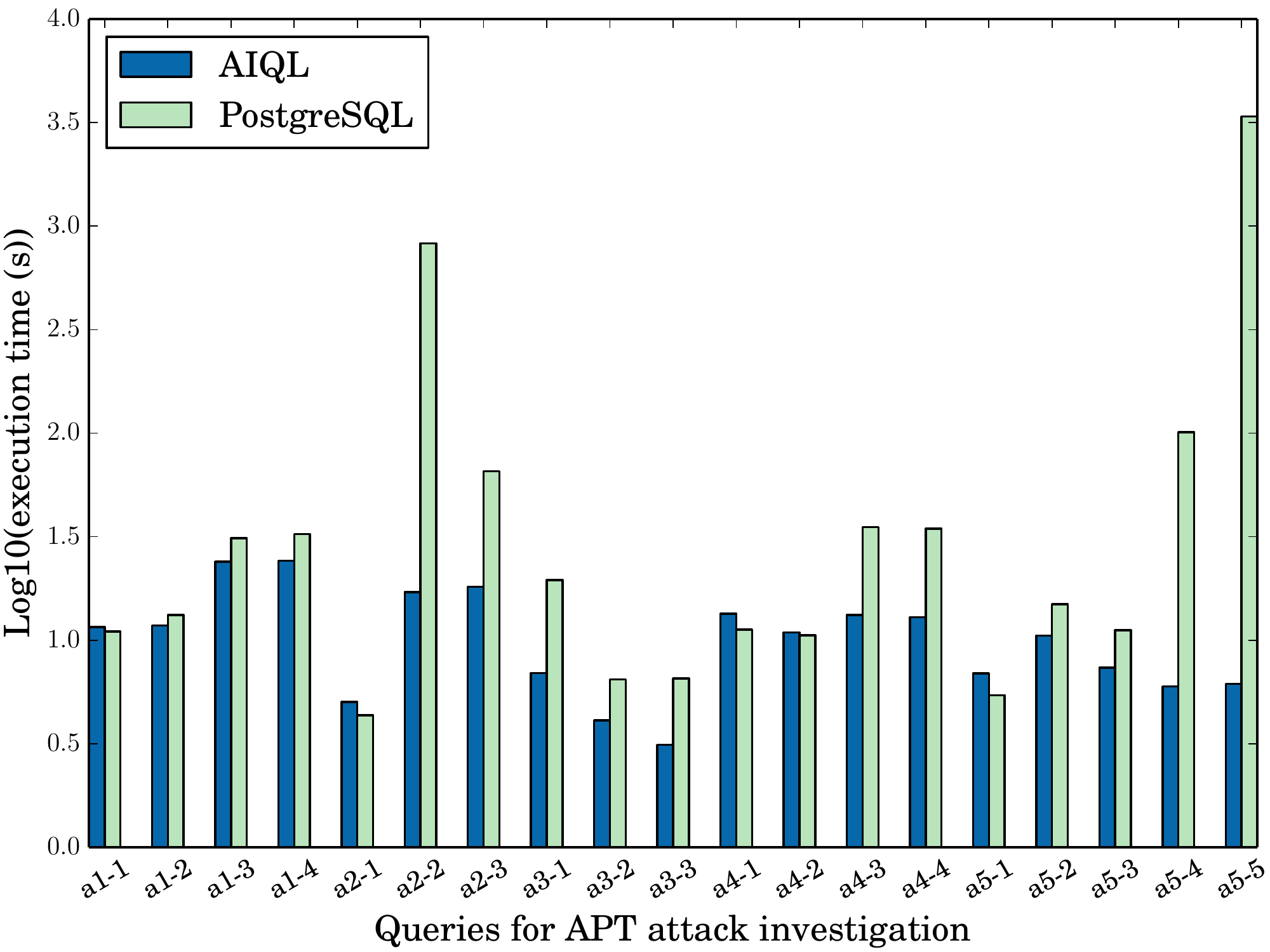}
\caption{Log10-transformed query execution time}
\label{fig:case-study-results}
\end{figure}

\myparatight{Post-Demo Evaluation: AIQL v.s. PostgreSQL (w/ Our Optimized Storage)}
Our investigation used 19 multievent queries and 1 anomaly query, touching 85 GB of data/257 million events.
\cref{fig:case-study-results} shows the log10-transformed execution time of \dsl queries and the semantically equivalent SQL queries executed in PostgreSQL.
Note that both \dsl and PostgreSQL employ our data storage optimizations.  
We observe clear superiority of \dsl in scheduling the execution of complex queries (e.g., \emph{a2-2, a5-5}).
The total execution time of \dsl is 3.6 minutes, achieving 21x performance speedup over PostgreSQL (77 minutes).
%
%
%
For the query conciseness, SQL queries contain at least 3.0x more constraints, 3.5x more words, and 5.2x more characters (excluding spaces) than AIQL queries.


\eat{
\begin{table}[!tp]
	\centering
	\caption{Aggregate statistics for case study}\label{tab:case}
	\begin{adjustbox}{width=0.44\textwidth}
		\begin{tabular}{|l|l|l|l|l|l|}
			\hline
			Attack Step &$\#$ of Queries 	& $\#$  of Evt Patterns	& \dsl (s)	& PostgreSQL (s)	& Neo4j (s)	\\\hline
			c1 	&1		&3			&3.8			&3.1				& 10.8		\\\hline
			c2 	&8		&27			&31.0		&8038.7			& 10981.7		\\\hline
			c3 	&2		&4			&15.9		&15.3 			& 3615.6		\\\hline
			c4 	&8		&35			&61.0		&10906.7			& 8150.6	 	\\\hline
			c5 	&7		&18			&58.8		&2166.5 			& 4285.4		\\\hline
			All 	&26		&87			&170.5		&21130.3 			& 27044.1		\\\hline
		\end{tabular}
	\end{adjustbox}
\end{table}
}

\begin{figure}[!tp]
\centering
\includegraphics[width=0.41\textwidth]{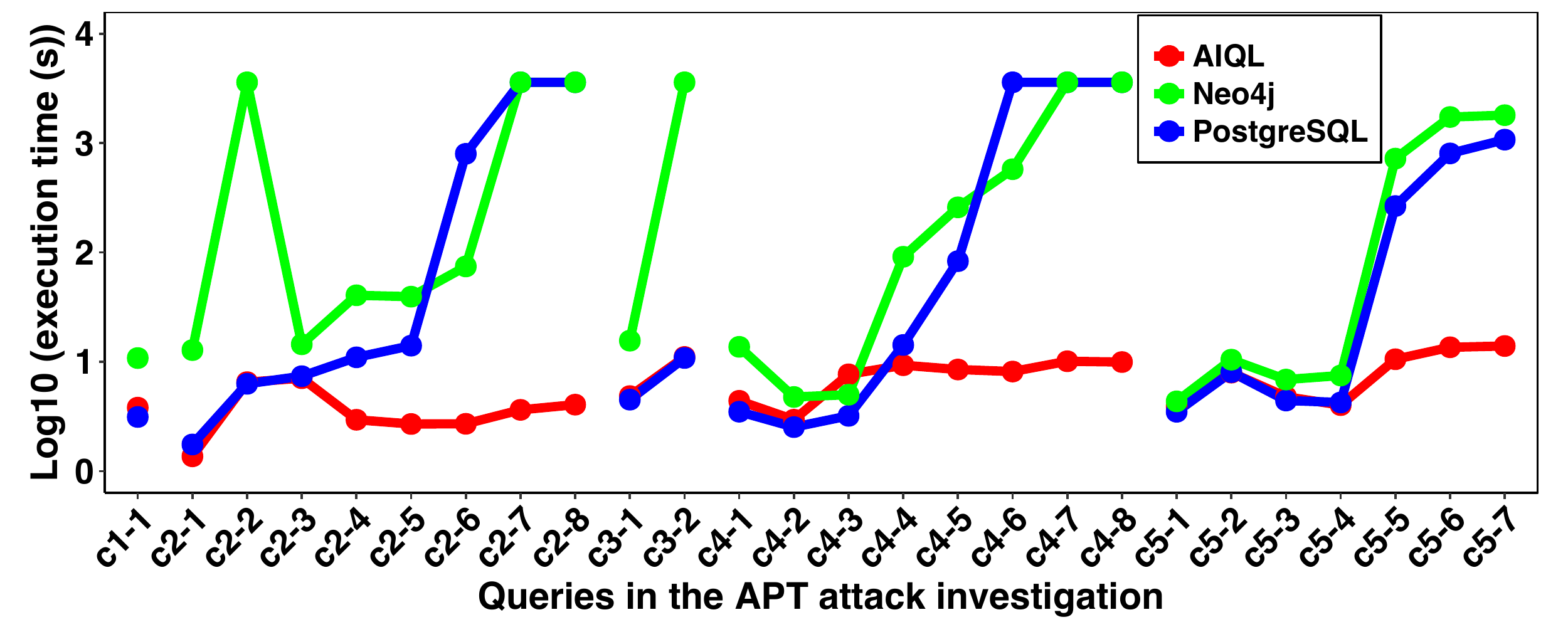}
\caption{Log10-transformed query execution time for another APT attack in~\protect\cite{aiql}}
\label{fig:case-study-results-outlook}
\end{figure}

\myparatight{Post-Demo Evaluation: AIQL v.s. PostgreSQL (w/o Our Optimized Storage) v.s. Neo4j}
In another case study of APT attack~\cite{aiql}, we evaluated the performance of \dsl against PostgreSQL w/o our optimizations and Neo4j.
As shown in \cref{fig:case-study-results-outlook}, the \dsl system as a whole is much faster than PostgreSQL (124x speedup) and Neo4j (157x speedup). 
In particular, Neo4j runs generally slower than PostgreSQL since it lacks support for efficient joins, which are required in expressing attack behaviors with multiple steps.
As the attack behaviors become more complex, besides performance degradation,
both SQL and Cypher queries become quite verbose with many joins and constraints, 
making it labor-intensive and error prone in constructing queries for timely attack investigation~\cite{aiql}.
%

\balance



\bibliographystyle{abbrv}
\bibliography{ref}

%


\end{document}